\documentstyle[preprint,aps,prd]{revtex}

\begin{document}
\draft
\title{Optical approach for the thermal partition function of
photons}
\author{Valter Moretti\footnote{Electronic address:
\sl moretti@science.unitn.it} 
and Devis Iellici\footnote{Electronic address:
\sl iellici@science.unitn.it}  }
\address{Dipartimento di Fisica, Universit\`a di Trento, \protect\\
and Istituto Nazionale di Fisica Nucleare,\protect\\
Gruppo Collegato di Trento \protect\\ I-38050 Povo (TN), Italy}

\date{October 1996}
\preprint{Preprint UTF-388}

\maketitle

\begin{abstract}

{\small   The optical manifold method to compute the one-loop
effective action in a static space-time is extended from the massless
scalar field to the Maxwell field in any Feynman-like covariant gauge.
The method is applied to the case of the Rindler space obtaining the
same results as the point-splitting procedure. The result is free from
Kabat's surface terms which instead affect the $\zeta$-function or
heat-kernel approaches working directly in the static manifold
containing conical singularities. The relation between the optical
method and the direct $\zeta$-function approach on the Euclidean
Rindler manifold is discussed both in the scalar and the photon cases.
Problems with the thermodynamic self-consistency of the  results
obtained from the stress tensor in the case of the Rindler space are
pointed out.}

\end{abstract}


\narrowtext

\section*{Introduction}

In a recent paper, \cite{ielmo}, we have computed the one-loop thermal
partition function of photons in the Rindler wedge employing a local
$\zeta$-function method directly in the Euclidean Rindler space.
Although this approach produces thermodynamical quantities with the
correct high temperature behaviour requested by the  statistical
mechanics, the low temperature behaviour seems to remain different
from that obtained with others methods. This can be seen
by means of a direct comparison between the free energy following 
from the above cited approach and the same quantity obtained by the
point-splitting renormalization procedure for the stress tensor
\cite{DC79,D087b,DO94}. In particular, one sees that the direct
$\zeta$-function approach gives, for the coefficient of the term
proportional to $T^2$, a result which is one-third of the
point-splitting result. This discrepancy can be traced back to an
identical discrepancy in the coefficients of the free energy of a
minimally coupled massless scalar field propagating in the Rindler
wedge \cite{ZCV,FRSE87,DO87a}.

It is important to remark that this problem does not arise from the
particular method used in \cite{ZCV} and \cite{ielmo} to compute the
determinant of the small fluctuations operator which appears in the
one-loop free energy. In fact, the same discrepancy has also been
found in \cite{CKV} using a completely different method to compute the
determinant. Therefore, it seems to be intrinsic of the computations
made directly in the Euclidean Rindler space.

In the photon and graviton case, a further drawback of the approach in
\cite{ielmo} is the need of a more complicated regularization
procedure  due to the presence of gauge depending ``surface'' terms
\cite{kabat}. Anyway, the results of \cite{ZCV,ielmo} improve
previous results obtained using global heat-kernel approaches
\cite{CW94,kabat} in the Rindler space, which is not able to reproduce
the Planckian high temperature behaviour.

There is another method which can be used to compute these one-loop
quantities, and is the optical one
\cite{GIPER78,BROT85,GIDD94,BARB94,EMP95,CVZ95,BAEMP95,dAO}. 
In this approach, instead of computing the partition function directly
in the static metric, one performs a conformal transformation in such
a way that the resulting manifold has an ultrastatic metric. Then, one
can compute the relevant quantities in this ``optical manifold'' using
heat-kernel, $\zeta$-function, or any other method and taking into
account how the various quantities transform under conformal
transformations. This method is particularly favorable in the
Euclidean Rindler case, since this manifold has a conical singularity
which can be quite tricky to deal with, whereas the related optical
manifold has no singularity. However, there is more in this method
than the mathematical content. In fact, it has been shown
\cite{GIDD94,BARB94,BAEMP95,dAO} that the canonical partition
function of a quantum field in a curved background with a static
metric is not directly related to the Euclidean path integral with
periodic imaginary time in the static manifold, but rather it is equal
to Euclidean path integral in the related optical manifold. In
particular, in \cite{BARB94} it is shown that the statistical counting
of states leads naturally to a formulation in the optical manifold. We
can also notice that, as far as we know, the equivalence of the direct
periodic imaginary time path integral formalism to the canonical
formalism for computing finite temperature effects has been proved in
ultrastatic manifolds only \cite{ALLEN86}.

Therefore, the computation of the thermal partition function from
Euclidean path integral in the static manifold requires the knowledge
of the Jacobian of the conformal transformation. On regular manifolds
this causes no trouble, since it is easily shown (see, e.g.,
\cite{report}) that the Jacobian affects only the
temperature-independent part of the free energy. Instead, as we will
see, when in the static metric there is a conical singularity the
temperature dependence of the Jacobian could be less trivial and
affect the temperature-dependent part of the free energy.

Another important point is that the optical method produces
thermodynamic quantities which agree with those obtained form the
point-splitting procedure. This happens in the case of a massless
scalar field  conformally coupled in the Rindler wedge at least, but
also, as we shall see, in the case of the photon field.

In the first part of this paper we shall review the computation of the
thermodynamical quantities of a massless scalar field in the Rindler
wedge, comparing the point-splitting, local $\zeta$-function and
optical results. In particular, we  note that while the
point-splitting approach can be applied for any coupling of the scalar
field with the gravity, the optical approach is feasible only for the
conformally coupled case, where it gives the same result as the
point-splitting method. Moreover, the dependence on the coupling
parameter, which disappears in the integrated physical quantities when the
background is a regular manifold, in this case affects these
quantities because of the presence of the conical singularity. On the
other hand, the computations made directly in the static Euclidean
Rindler manifold using the local $\zeta$-function technique is still
limited to the minimally coupled case, and even in this case the
result is different from the point-splitting one. 

In the second part of this work, we shall extend the optical
manifold approach to the Maxwell field case. We shall show that 
 there are two possible ways to do this, which are equivalent in the
scalar case, but in the photon case could produce a different result.
The difference of the two approaches is essentially in the definition
of the gauge-fixing and ghost parts of the Lagrangian. In particular,
we shall show that result expected counting the polarization states 
of the field can be obtained by defining the
theory directly in the optical manifold. The other possibility is to
define the partition function in the Euclidean Rindler wedge and only
then perform the conformal transformation.  We are not able to work
out thoroughly this latter approach, because of a mathematical
complication in the ghost action. Nevertheless, we argue that it gives
a different coefficient of the term proportional to $T^2$.

In the final part, in addition to  discussing the obtained results,
we also show that the usual relations between
thermodynamical quantities, which involve derivatives with 
respect to the temperature, lead to inconsistencies when 
applied to Rindler thermal states with temperature different
from the Unruh-Hawking one.

\section{The general problem in the case of  a massless scalar field}

As we have said in the Introduction, the problem of discrepancy of the
coefficient of $T^2$ in the free energy appears already in the case of
a massless scalar field propagating in the Rindler wedge. In fact,
this problem seems to be independent of the field spin. Hence, we
start discussing just this case, taking into consideration the
point-splitting, local $\zeta$-function, and optical manifold
approaches.

\subsection{Point-splitting approach}

We start considering the results  produced by the point-splitting
method. The point-splitting renormalized stress tensor reads (see,
for example \cite{FRSE87}, continuing into the Rindler space the
results obtained for the cosmic string)
\begin{eqnarray}
\langle T^\mu_\nu\rangle_\beta^{\scriptsize \mbox{p.-s.}}(\xi) &=&
\frac{1}{1440 \pi^2 r^4} \left[\left(
\left(\frac{2\pi}{\beta}\right)^4 -1\right) \mbox{diag}(-3,1,1,1) 
\right.
\nonumber\\
&&\left. +20(6\xi-1)\left( \left(\frac{2\pi}{\beta}\right)^2 -1\right)
\mbox{diag}(\frac{3}{2},-\frac{1}{2},1,1) \right] \label{tensor}
\end{eqnarray}
By integrating $-\sqrt{g} T^0_0$, we get a total energy which we shall
compare with those following from the other methods:
\begin{eqnarray}
U_{\beta,\xi}^{\scriptsize \mbox{p.-s.}} =
\frac{L_y L_z}{2880\pi^2\epsilon^2}\left[ 
3\left(\frac{2\pi}{\beta}\right)^4 -
 30 (6\xi -1) \left(\frac{2\pi}{\beta}\right)^2 + 30 (6\xi -1) -3 
\right] 
 \label{i3}\:.
\end{eqnarray}
Above, $L_y$ and $L_z$ are the (infinite) lengths of the transverse
dimensions, and so $ A_\perp = L_y L_z$ is the (infinite) area of the
horizon. The parameter $\xi$ fixes the coupling of the scalar field
with the gravitation. In the (Lorentzian) Rindler space the scalar
curvature $R$ is zero everywhere and the parameter $\xi$ remains as a
relic of the fact that $T_{\mu\nu}$ is obtained by varying the metric
$g_{\mu\nu}$ in the field Lagrangian \cite{birrel}. \footnote{It is
worthwhile noticing that one has to consider the theory within the
curved space time in order to discuss on the physics in the flat
space-time. Anyhow, the extension of the theory to a curved space-time is
not unique  and this involves some subtleties regarding also the
regularization procedure. The choice between different regularization
procedures should be made on the basis of what  is the physics that
one is trying to describe. Obviously, the general hope is that, at the end of 
the complete renormalization procedure involving matter fields and gravity,
all these different regularization approaches give rise to equivalent
physical results.  See \cite{BAEMP95} and \cite{LAWI96} for a
discussion on these topics.}
 Employing the general expression of $T_{\mu\nu}(\xi)$
\cite{FRSE87,birrel} in terms of the Hadamard function, one finds
that, in the case $R=0$, the global conserved quantities as total
energy should not depend on the value of $\xi$. This is  because the
contributions to those quantities due to $\xi$ are discarded into
boundary surface integrals which generally vanish. However, this is
not the case dealing with the Rindler wedge because such integrals
diverge therein.\footnote{Similar problems appear working in
subregions of the Minkowski space in the presence of boundary conditions
\cite{birrel}.}
 The
only possibility to get a result not depending on $\xi$ consists in
taking $\beta=2\pi$ producing a trivial result. The considered
ambiguity does not seem to arise from a similar ambiguity in defining
the thermal quantum state. In fact, the thermal Wightman functions
employed in calculating the renormalized stress tensor do not depend
on $\xi$. Finally, it is worthwhile noting that the $\xi$ ambiguity
affects the $\beta^{-2}$ term in the thermodynamical quantities and
hence their low temperature behaviour.

Notice that Kay and Studer \cite{KS91} found an ambiguity in defining
the scalar Wightman functions around a cosmic string, a background
which has the same Wick-rotated metric as the thermal Euclidean
Rindler manifold. However, this ambiguity is related to the
time-independent modes and so, e. g., employing the $\zeta$-function
procedure, one simply finds that this ambiguity cannot produce
$\beta^{-2}$ terms in the Rindler free energy. Hence, it should not be
related with the $\xi$-ambiguity.

\subsection{Direct conical approach}

One can formally define the partition function at the temperature
$1/\beta$ by an Euclidean path integral
\begin{eqnarray}
e^{-\beta F_\beta} = Z_\beta = \int {\cal D} \phi \exp \left(-S[\phi]\right)\:,
\label{i1}
\end{eqnarray}
where the (Euclidean) action is that of a massless scalar field
coupled with the gravitation,
\begin{eqnarray}
S[\phi] = - \frac{1}{2}\int d^4x \sqrt{g}\:\phi
\left[\nabla_\mu\nabla^\mu -\xi R\right]\phi.
\label{i1.5}
\end{eqnarray}
The background is the Euclidean Rindler manifold ${\cal
C}_\beta\times R^2$ with an imaginary time period $\beta$. The 
Euclidean Rindler  metric reads
\begin{eqnarray}
ds^2 = r^2d\theta^2 + dr^2 + dy^2 +dz^2 \label{rindler}\:,
\end{eqnarray}
where $\theta \in [0 , \beta]$, $r\in R^+$, ${\bf x}= (y,z) \in R^2$.
Notice the well-known conical singularity at $r=0$ when
$\beta\neq 2\pi$.

In the case $\xi=0$, the previous partition function  can be
explicitly computed by a local $\zeta$-function approach recently
introduced by Zerbini {\em et al.} \cite{ZCV} obtaining a Minkowski
renormalized free energy $F_\beta^{\scriptsize \mbox{sub}} = F_\beta
-U_{\beta=2\pi}$ and a renormalized internal energy\footnote{As it is
well known, the $(\beta=2\pi)$-thermal Rindler state locally coincides
with the Minkowski vacuum and, in renormalizing, we suppose that this
state does not carries energy density. Notice that such a Minkowski
subtraction procedure does not affect the entropy computed through
$F_\beta$.} $U_\beta^{\scriptsize\mbox{sub}} = \partial_\beta \beta
F_\beta - (\partial_\beta \beta F_\beta)|_{\beta=2\pi} $ which read
\begin{eqnarray}
F_\beta^{\scriptsize \mbox{sub}} &=&-\frac{A_\perp}{2880 \pi^2
\epsilon^2} \left[  \left( \frac{2\pi}{\beta}\right)^4 +10  \left(
\frac{2\pi}{\beta}\right)^2 +13 \right] \label{f},\nonumber\\
U_\beta^{\scriptsize\mbox{sub}} &=& \frac{A_\perp}{2880 \pi^2
\epsilon^2} \left[3\left( \frac{2\pi}{\beta}\right)^4
+10  \left(\frac{2\pi}{\beta}\right)^2 -13\right]
\label{FU},
\end{eqnarray}
where $A_\perp$ is the (infinite) event horizon  area and $\epsilon$ a
short-distance cutoff representing the minimal distance from the
horizon \cite{thooft}. We note that the above result is different from
that of the point-splitting with $\xi=0$ and the difference is in the
coefficient of the term proportional to $\beta^{-2}$. However, as we
said in the Introduction, the partition function (\ref{i1}) is related
to the canonical one by the Jacobian of a conformal transformation:
this could explain the different result. We shall come back on this
point after discussing the optical approach.

It is worthwhile noticing that the Lorentz section of the Rindler
space is flat and hence, as far as the real time theory is concerned,
we find a complete independence on the parameter $\xi$. However, in
calculating the partition function, one has to deal with the Euclidean
section of the Rindler manifold and, considering it as a integral
kernel, the curvature $R$ takes  Dirac's $\delta$ behaviour at $r=0$
\cite{SOL95,FUSOL95}, thus the value of the parameter $\xi$ could be
important. The previous results have been carried out in the case
$\xi=0$ in the sense that the eigenfunctions employed in computing the
$\zeta$ functions properly satisfy the eigenvalue equation with no $R$
term.

In the case $\xi\neq 0$ the problems are due to the fact that the
equation for the eigenfunctions contains a Dirac $\delta$, and so it is
not mathematically clear how to treat it. In the case of a cosmic
string, the Dirac $\delta$ represents a limit case, maybe unphysical, of
the problem in which the string has a finite thickness, which is
mathematically well defined since no Dirac $\delta$ appears. In the case
of the Rindler space there is no such way out, and the only way to 
avoid the problem is to consider the case $\xi=0$.

\subsection{Optical approach}

Let us now consider the optical approach
\cite{GIPER78,BROT85,GIDD94,BARB94,EMP95,CVZ95,BAEMP95,dAO}.
As remarked in the Introduction, this approach is not just a
mathematical method to compute the functional integral in Eq.
(\ref{i1}), but has an important physical content. In fact, as
previously remarked, it fulfills the requirements of a formulation in
the optical manifold following from statistical counting of states.

Let us consider a static metric $ds^2 = g_{\mu\nu}dx^\mu dx^\nu$ and
perform a conformal transformation of the metric 
(maybe singular if $\Omega(x)=0$), $g_{\mu\nu}\rightarrow
g'_{\mu\nu}=\Omega^{2}(x)g_{\mu\nu}$, so that
\begin{eqnarray}
ds^2 \rightarrow ds'^2= \Omega^{2} ds^2 \:.
\label{conformal0}
\end{eqnarray}
Choosing $\Omega^{-2} = g_{00}$, $ds'^2$ becomes the related ultrastatic
optical metric. In the case of the Euclidean Rindler space, this
conformal factor becomes singular just on the conical singularities,
which are pushed away to the infinity\footnote{The points at $r=0$ of
the optical manifold $S^1\times H^3$ are infinitely far from the
points of the manifold with $r>0$ taking the distance  as the affine
parameter  along geodesics. Strictly speaking, the former  points do
not belong to the manifold at all.} and the optical manifold is free
from singularities. Under such a transformation, the massless scalar
field $\phi$ transforms into $\phi' = \Omega^{-1} \phi$ and the
Euclidean action with coupling factor $\xi$ transforms into
the following more complicated action \cite{birrel}:
\begin{eqnarray}
S'[\phi'] =  - \frac{1}{2}\int d^4x \sqrt{g'}\: \phi' \left[
\nabla'_\mu  \nabla'^\mu - \xi_4 R'-\Omega^{-2}(\xi-\xi_4) R\right]
\phi'
 \label{i1.5'}\:,
\end{eqnarray}  
where $\xi_D=(D-2)/4(D-1)$ is the conformal invariant factor. If we
consider a conformally coupled field, $\xi=\xi_D$, we see that also
the transformed action is that of a conformally coupled field in the
optical manifold $S^1\times H^3$. In the other cases, we have to keep
a term proportional to $R$ which has a Dirac $\delta$ behaviour at
$r=0$ and thus we have an ill-defined operator.\footnote{One possible
way to get rid of this term is to define the action in the Lorentzian
manifold, where $R=0$, perform the conformal transformation to the
optical manifold, and only then use the transformed action to write to
partition function with the periodic imaginary time formalism
\cite{EMP95}. This procedure gives a result independent on the
parameter $\xi$ by nature: the coupling in the optical manifold is
always conformal. However, in our opinion this procedure seems too
{\em ad hoc}.}

When we compute the one-loop partition function, we may formally
write \cite{POL81,DS89}
\begin{eqnarray}
e^{-\beta F_\beta} = 
 Z_\beta = \int {\cal D} \phi' \: J[g,g',\beta] \: \exp -S'[\phi'] =
J[g,g',\beta] Z'_\beta = J[g,g',\beta] e^{-\beta F'_\beta}\:. 
 \label{foundamental}
\end{eqnarray}
We remark that is $Z'_\beta$ which is equivalent to the canonical
partition function \cite{GIDD94,BARB94,BAEMP95,dAO}.
The functional Jacobian $J[g,g',\beta]$ does not depend on $\phi'$ and
thus it can be carried out from the integral as we have done above.
When the involved manifolds are regular, it is possible to prove that
such a Jacobian is in the form
\begin{eqnarray}
J[g,g',\beta] = \exp (-\beta E_0) \label{jwb} \:.
\label{jaco}
\end{eqnarray}
where  $E_0$ does not depend on $\beta$. This is substantially due to
the staticity of the involved metrics \cite{CVZ95,BAEMP95,dAO} and the
factor $\beta$ in the exponent is due to an  integration over the
whole Euclidean manifold. If this holds in the presence of a conical
singularity as well, one  expects that $F_\beta$ and $F'_\beta$ differ
only for the value of the renormalized zero-temperature energy. When
the coupling in the Euclidean Rindler manifold is conformal, the
direct computation of $F'_\beta$ can be performed employing the
$\zeta$-function approach \cite{CVZ95} (see also the Appendix of this
paper). We report here the well-known final result only:
\begin{eqnarray}
F'_\beta = - \frac{A_\perp}{2880\pi^2\epsilon^2}\left( \frac{2\pi}{\beta}
 \right)^4\:. \label{f'}
\end{eqnarray}
Using $U'_\beta=\partial_\beta(\beta F'_\beta)$ to compute the
internal energy and performing the Minkowski renormalization
${U'_\beta}^{\scriptsize \mbox{sub}} = U'_\beta - U'_{\beta=2\pi}$ in
order to get a vanishing internal energy at $\beta= 2\pi $,  we find
just
\begin{eqnarray}
{U'_\beta}^{\scriptsize \mbox{sub}}
= U_{\beta, \xi=1/6}^{\scriptsize \mbox{p.-s.}}\:.
\end{eqnarray}
Therefore, we have got a result equal to the point-splitting one 
by renormalizing (with respect to the Minkowski vacuum)  the internal
energy obtained on the optical manifold and {\em without} taking into
account the Jacobian, whether it has the form (\ref{jaco}) or not.

\subsection{Comparison of the results}

In the previous subsection we have seen that the optical method, when
applicable, gives the same result as the point splitting. On the other
hand, we see that $U_{\beta,\xi}^{\scriptsize \mbox{p.-s.}}$ does not
coincide with the corresponding internal energy (\ref{FU}) found by
the $\zeta$-function approach at the value of coupling parameter one
expects, $\xi=0$, but rather at $\xi=1/9$. Note that the discrepancy
is in the term proportional to $\beta^{-2}$, while the difference in
the $\beta$-independent term is not meaningful, because such terms 
are fixed by the subtraction procedure: they do coincide when the
remaining terms are equal. Note also that we cannot compare directly
the optical and the local $\zeta$-function approaches, since they are
not applicable for the same value of $\xi$.

In order to identify the source of the above discrepancy, we remind
the reader
that the thermodynamical quantities in Eq. (\ref{FU}) have been
obtained from the Euclidean path integral in the static manifold,
which differs from the canonical partition function for the Jacobian
of the conformal transformation, see Eq. (\ref{foundamental}). As we
have said above, on regular manifolds the logarithm of this Jacobian
is simply proportional to $\beta$, thus giving a contribution only to
the temperature-independent part of the free energy. However, the case
of the Euclidean Rindler space could be more complicated, due to the
presence of a conical singularity at $r=0$, which could yield a
nonlinear dependence on $\beta$. In fact, such a singularity can be
represented as an opportune Dirac $\delta$ function with a coefficient
containing a factor $(2\pi-\beta)$ \cite{SOL95,FUSOL95}, and so
$\beta$ enters not only as integration interval, but also in the
integrand. Of course, only an explicit calculation of the Jacobian can
give an ultimate answer. In two dimensions, the Jacobian
$J[g,g',\beta]$ is the exponential of the well-known Liouville action
\cite{POL81}, and an easy calculation shows that the logarithm of the
Jacobian is indeed proportional to $\beta$ \cite{dAO}, regardless of
the conical singularity. Unfortunately, in four dimensions the form of
the Jacobian is far more complicated (see \cite{dAO}  and references
therein) and involves also products of curvature tensors which are ill
defined. Therefore, it is not clear whether the discrepancy in the
term proportional to $\beta^{-2}$ might be assigned to the Jacobian.

Summarizing, we have seen that the optical method has been applied  
to the conformally coupled case only, and in this case it gives
the same result as the point-splitting method. With regard of
the direct computation in the Euclidean Rindler wedge, it has to be
considered as incomplete, because of our ignorance of the Jacobian 
and of the nonminimally coupled case. We stress that, in the case of
a regular manifold, these two approaches should be equivalent.

\section{Optical approaches in the case of photons}

In \cite{ielmo} the partition function of photons gas in a Rindler
wedge has been computed generalizing the procedure in \cite{ZCV}. The
found Minkowski renormalized free energy  amounts to $2 F^{\scriptsize
\mbox{sub}}_\beta + (2 - \ln \alpha) F_\beta^{\scriptsize
\mbox{surface}}$, where $F_\beta^{\scriptsize \mbox{sub}}$ is the
scalar free energy  previously discussed, Eq. (\ref{FU}), and the
$F_\beta^{\scriptsize \mbox{surface}}$ is a ``surface'' term which
arises integrating a total derivative and has the form $A_\perp
[(2\pi/\beta)^2-1]/(24\pi^2\epsilon^2)$ (see \cite{kabat} and
\cite{ielmo} for more comments), finally $\alpha$ is the gauge-fixing
parameter. Notice that also this anomalous gauge-dependent term
involves a $\beta^{-2}$ dependence. We suggested  droping this latter
gauge-dependent term as the simplest procedure to remove the
unphysical gauge dependence. Anyway, we stressed that other procedures
could also be possible. The obtained result agrees with the
statistical mechanics request at high temperatures, but, as in the
scalar case, the low temperature behaviour is different from that
obtained with the usual point-splitting procedure. Therefore, let us
consider the point-splitting results \cite{DC79,FRSE87}: the
renormalized stress tensor takes a simple form
\begin{eqnarray}
\langle T^\mu_\nu\rangle_\beta^{\scriptsize \mbox{phot. p.-s.}} = 
\frac{1}{720\pi^2 r^4}\left [ \left( \frac{2\pi}{\beta}\right)^4 
+ 10 \left( \frac{2\pi}{\beta}\right)^2 - 11 \right]\,
\mbox{diag}(-3,1,1,1).
\label{tensor1}
\end{eqnarray}
The (Minkowski renormalized) internal energy
corresponding to the previous photon stress tensor reads
\begin{eqnarray}
U_\beta^{\scriptsize \mbox{phot. p.-s.}} 
= \frac{3A_\perp}{1440 \pi^2 \epsilon^2}\left[
 \left( \frac{2\pi}{\beta}\right)^4 
+ 10 \left( \frac{2\pi}{\beta}\right)^2 -11 \right].
\end{eqnarray}
As far as the energy density is concerned, we have the following 
very simple relation:
\begin{eqnarray}
\langle T^0_0\rangle_\beta^{\scriptsize \mbox{phot. p.-s.}} =
2 \langle T^0_0\rangle_\beta^{\scriptsize \mbox{p.-s.}}(\xi = 0) \:,
\label{tensor2}
\end{eqnarray}
where on the right the stress tensor is that of a massless scalar
field. It is worth while noticing that $\xi = 0$ takes place on the
right-hand side instead of $\xi= 1/6$. Hence, the energy density of
the electromagnetic field does not amount to twice that of a
conformally coupled scalar field, as one could naively expect
considering that the electromagnetic field is conformal invariant in
four dimensions. As far as the internal energy is concerned, we find
the same unforeseen relation. However, as previously discussed, the
integrated quantities should  not have to depend on $\xi$  in  more
``regular'' theories,  restoring the naively expected relation
between the considered quantities.

Discussing the scalar case we have stressed the importance of
the optical method: therefore, now we go to investigate 
whether it is possible or not to get such an energy employing the
optical-manifold method. There are two possible ways to implement 
this method. The simplest one consists of defining the partition
function directly as a functional integral on the optical manifold.
However, there is another more complicated possibility: it consists of
starting with a functional integral in the initial static manifold,
performing the conformal transformation and finally dropping the
functional Jacobian. This is, in fact, the simplest generalization of
the results obtained in the scalar case. Both  methods produce the
same final functional integral in the simpler conformally coupled
scalar case, but in the case of the Maxwell field the two procedures
do not seem to be equivalent, as we shall see,  due to the presence of
gauge-fixing and ghost terms.

\subsection{Optical approaches in the case of  general static
manifolds}

Let us start reviewing the formalism we use dealing with the photon
field. The complete action for the electromagnetic field in any
covariant gauge and on a general Euclidean manifold, endowed with a
metric $ds^2 = g_{\mu\nu} dx^\mu dx^\nu$, which we shall suppose
{\em static} and where $\partial_0$ is the global (Euclidean) timelike
Killing vector with closed orbits of period $\beta$. Using the in
Hodge de Rham formalism  we have
\begin{eqnarray}
S^{\scriptsize \mbox{em}}&=&
\int d^4x\left[\frac{1}{4}  \langle F,F\rangle + \frac{1}{2\alpha}
\langle A,d\delta A\rangle\right] + S_{\scriptsize
\mbox{ghost}}(\alpha)\nonumber\\
&=& \frac{1}{2}\int d^4x\left[\langle  A,\Delta A\rangle - 
\left(1-\frac{1}{\alpha}\right)\langle A,d\delta A\rangle\right]+ 
S_{\scriptsize \mbox{ghost}}(\alpha)
\label{1}\:.
\end{eqnarray}
In order to maintain the gauge invariance of the theory, it is
important to keep the dependence on the gauge-fixing parameter of the
ghost action, as one obtains by varying the gauge-fixing condition
$\frac{1}{\sqrt{\alpha}}\delta A=0$ \cite{NvN}:
\begin{eqnarray}
S_{\scriptsize \mbox{ghost}}(\alpha) = -\frac{1}{\sqrt{\alpha}}\int
d^4x \sqrt{g}\: \overline{c}\Delta c,
\label{ghost}
\end{eqnarray}
where $\Delta$ is the Hodge-de Rham Laplacian for 0-forms and $c$,
$\overline{c}$ are anticommuting scalar fields.
Usually, the dependence on the gauge-fixing parameter is absorbed
rescaling the ghost fields, but in the presence of a scale anomaly this
rescaling gives rise to a nontrivial contribution, which is essential
to maintain the gauge invariance of the theory. This is just the case
here: in fact, the contribution of the action (\ref{ghost}) to the
one-loop effective action is proportional to that of a minimally
coupled scalar field, which has a scale anomaly in four dimensions.

Some comments on  the formalism in Eq. (\ref{1}) are in order.
$F\equiv \partial_\mu A_\nu -\partial_\nu A_\mu = 
\nabla_\mu A_\nu -\nabla_\nu A_\mu $ is the 2-form representing the
photon strength field, $\nabla_\mu$ being the covariant derivative;
the brackets stand for the $p$-forms Hodge local product:
\begin{eqnarray}
\langle G,H\rangle = G \wedge * H =\sqrt{g}\,
g^{\mu_1\nu_1}...g^{\mu_p\nu_p} G_{\mu_1... \mu_p} H_{\nu_1...\nu_p}
\nonumber
\end{eqnarray}
For future reference we also define the internal product
\begin{eqnarray}
G \cdot H = g^{\mu_1\nu_1}...g^{\mu_p\nu_p} G_{\mu_1... \mu_p}
H_{\nu_1...\nu_p}  \nonumber\:.
\end{eqnarray}
We remind the reader 
that $\delta = (-1)^{N(p+1)+1}*d*$ is the formal adjoint of
the operator $d$ with respect to the scalar product of $p$-forms
induced by the integration of the previous Hodge local product;
finally, $\Delta = d \delta + \delta d$ is the Hodge-de Rham Laplacian
of the $p$-forms. In order to perform calculations through the usual
covariant derivative formalism the following relations for $0$-forms
and $1$-forms are quite useful:
\begin{eqnarray}
\Delta \phi &=& -\nabla_\mu \nabla^\mu \phi \:, \nonumber\\ 
\delta A
&=& -\nabla_\mu A^\mu \nonumber\:,\\ (\Delta A)_\mu 
&=& - \nabla_\nu
\nabla^\nu A_\mu + R^\nu_\mu A_\nu\:.\nonumber
\end{eqnarray}

The second line of Eq. (\ref{1}) represents the complete photon action
now expressed in terms of the vector field $A_\mu$  and the ghost
fields only and it is the one usually employed in order to compute the
partition function of the photon field by means of a functional
integral. The partition function of photon at the temperature
$T=1/\beta$ is then formally expressed by 
\begin{eqnarray}
Z_\beta &=& \int {\cal D}A \: \exp
- \frac{1}{2}\int d^4x \,\left[\langle  A,\Delta A\rangle -
\left(1- \frac{1}{\alpha}\right) \langle A,d\delta A\rangle\right]
\nonumber \\
&& \times \int {\cal D}c {\cal D}\overline{c}
\exp  -S_{\scriptsize \mbox{ghost}}(\alpha) \nonumber\:.
\end{eqnarray}

In order to compute this partition function, we want to pass to the
related optical manifold, and so we consider a conformal
transformation, Eq. (\ref{conformal0}), with $\Omega^2 = g_{00}$.
Notice that, since we work in four dimensions, the $p$-forms $A$ and
$F$ have a vanishing mass dimension and thus they must be conformally
invariant, namely $A=A'$ and $F=F'$. Furthermore the following
identity arises:
\begin{eqnarray}
\langle F,F\rangle' &=& \langle F,F\rangle \label{2}\:.
\end{eqnarray}

\subsection{First general approach}

As we said above, the way to proceed is twofold. As a first way, we
can suppose to have performed the conformal transformation {\em
before} we start with the field theory. This means that we define the
partition function of photons in the static manifold as a path
integral directly in the optical manifold. In such a case the
expression of the
partition function is defined by
\begin{eqnarray}
Z^{(1)}_\beta &=& \int {\cal D}A \: \exp
- \frac{1}{2}\int d^4x\,\left[\langle  A,\Delta' A\rangle' -
\left(1- \frac{1}{\alpha}\right)\langle A,d\delta' A\rangle'\right]
\nonumber \\
&& \times \int {\cal D}c' {\cal D}\overline{c}' \exp
-S'^{(1)}_{\scriptsize \mbox{ghost}}(\alpha)
 \label{firstcase}\:,
\end{eqnarray}
where
\begin{eqnarray}
S'^{(1)}_{\scriptsize \mbox{ghost}}(\alpha) = -\frac{1}{\sqrt{\alpha}}
\int d^4x \sqrt{g'}\: \overline{c}'\Delta' c' \:,
\nonumber
\end{eqnarray}
and where the primed metric and variables appearing in the previous
functional integral are the optical ones. In other words, for the
one-loop Euclidean effective action $-\ln Z^{(1)}_\beta$ we have
\begin{eqnarray}
\ln Z^{(1)}_\beta = -\frac{1}{2} \ln \det \mu^{-2}\left[\Delta' - 
\left(1-\frac{1}{\alpha}\right)d\delta'\right]
+\ln Z^{(1)}_{\beta, \scriptsize \mbox{ghost}}(\alpha)
\label{zeta1'}\:.
\end{eqnarray}
Here $\mu$ is an arbitrary renormalization scale necessary on a
dimensional ground in the above formula and  denoting the presence of
a scale anomaly if it does not  disappear from the final formulae.

For future reference we  note that the effective action of the ghosts,
except for the $\alpha$ dependent factor, amounts trivially to minus
twice the Euclidean effective action of an uncharged massless scalar
field with the Euclidean action minimally coupled with the
gravitation. Therefore its contribution to the one-loop effective
action can be written immediately from the $\zeta$ function of a
minimally coupled scalar field,
$\zeta^{\scriptsize\mbox{m.c.s.}}(s;x)$, in the same background,
taking the $\alpha$-dependence into account:
\begin{eqnarray}
 \ln Z^{(1)}_{\beta, \scriptsize \mbox{ghost}}(\alpha)=
-\int d^4x\sqrt{g'}\left[\frac{d}{ds}\zeta^{\scriptsize
\mbox{m.c.s.}}(s;x)|_{s=0}+\zeta^{\scriptsize
\mbox{m.c.s.}}(s;x)|_{s=0}\ln \sqrt{\alpha}\mu^2\right].
\label{fantom1}
\end{eqnarray}

\subsection{Second general approach}

As a second way,  we can suppose to define the partition function
directly in the static manifold, adding also the gauge-fixing 
term and the ghost Lagrangian to the pure electromagnetic action,
and only {\em after} perform the conformal transformation to the
optical metric.
In this way we have to find how all the pieces in the path integral
transform under the conformal transformation. In particular, the
operator $\Delta - (1-\alpha^{-1}) d\delta$  transforms into another
operator $\Lambda_\alpha$, which we are going to write shortly. As
regards the functional Jacobian which arises from the functional
measure, a direct 
generalization of the discussion made in the scalar case
tells us that it has to be ignored if we are interested in computing
the thermal partition function. However, we would have to take it into
account if we were computing, for example, the zero-temperature
effective action in a cosmic string background.

Hence, employing this second procedure, we shall assume  the photon
partition function to be defined by
\begin{eqnarray}
Z^{(2)}_\beta &=&  \int {\cal D}A' \: \exp -\left\{\frac{1}{2}\int
d^4x \langle  A',\Lambda_\alpha A'\rangle'\right\}
 \int {\cal D}c' {\cal D}\overline{c'} \exp  -S'^{(2)}_{\scriptsize
\mbox{ghost}}(\alpha)\nonumber\:,
\end{eqnarray}
In other words, for the Euclidean effective action $-\ln
Z^{(2)}_\beta$ we have
\begin{eqnarray}
\ln Z^{(2)}_\beta =- \frac{1}{2} \ln \det (\mu^{-2} \Lambda_\alpha )
+\ln Z^{(2)}_{\beta, \scriptsize \mbox{ghost}}(\alpha)
\label{zeta2'}\:.
\end{eqnarray}
The form of $S'^{(2)}_{\scriptsize \mbox{ghost}}(\alpha)$
is that of Eq. (\ref{ghost}) after a conformal
transformation:
\begin{eqnarray}
S'^{(2)}_{\scriptsize \mbox{ghost}}(\alpha) = -
\frac{1}{\sqrt{\alpha}}\int d^4x \sqrt{g'} \: \overline{c}'
\left[\Delta' +\frac{1}{6}( R'-\Omega^{-2} R)\right]   c'
\label{5}\:,
\end{eqnarray}
where $c'=\Omega c $, $\overline{c}'=\Omega\overline{c}$. For future
reference, we note that this effective action of the ghosts amounts
trivially to minus twice the Euclidean effective action of an
uncharged massless scalar field $\varphi$ with the Euclidean action
($\Delta' = - \nabla'_\mu\nabla'^\mu$) endowed by an
$\alpha-$depending overall factor
\begin{eqnarray}
S^{(2)}(\alpha) =\frac{1}{\sqrt{\alpha}} \int d^4x \sqrt{g'} \:
\frac{1}{2}\varphi \left[\Delta' +\frac{1}{6}( R'-\Omega^{-2}
R)\right] \varphi
\label{6'}\:.
\label{ghost2}
\end{eqnarray}
When the static manifold is flat, $R=0$, the contribution of the
ghosts to the effective action can be written in terms of the $\zeta$
function of a conformally coupled scalar field:
\begin{eqnarray}
\ln Z^{(2)}_{\beta, \scriptsize \mbox{ghost}}(\alpha)=
-\int d^4x\sqrt{g'}\left [\frac{d}{ds}\zeta^{\scriptsize
\mbox{c.c.s.}}(s;x)|_{s=0}+\zeta^{\scriptsize
\mbox{c.c.s.}}(s;x)|_{s=0}\ln \sqrt{\alpha}\mu^2\right].
\label{fantom2}
\end{eqnarray}

Now, let us find the explicit form of the operator $\Lambda_\alpha$.
The following identity holds:
\begin{eqnarray}
\delta A &=& \frac{1}{\Omega} (\delta' A - \eta\cdot A) \label{2.5}\:,
\end{eqnarray}
provided  the $1$-form $\eta$ be defined as
\begin{eqnarray}
\eta = d (\ln \Omega) \equiv  \partial_\mu \ln \Omega \label{3}\:.
\end{eqnarray}
Taking into account that $\delta  = d^\dagger$ and employing Eq.s
(\ref{1}), (\ref{2}) and (\ref{3}) we get the identity
\begin{eqnarray}
S^{\scriptsize \mbox{em}}&=& \frac{1}{2}\int d^4x \left[\langle  A,
\Delta A\rangle - \left(1-\frac{1}{\alpha}\right) \langle  A, d\delta
A\rangle\right]+ S_{\scriptsize \mbox{ghost}}(\alpha)
\nonumber \\
&=& \frac{1}{2}\int d^4x\left[\langle  A, \Delta' A\rangle' -
\left(1-\frac{1}{\alpha}\right) \langle  A, d\delta' A\rangle'
+\frac{1}{\alpha}\langle A,\eta \eta\cdot A\rangle' \right.
\nonumber\\
&&\left. -\frac{1}{\alpha}\langle A, (\eta\delta' + d \eta \cdot)
A\rangle' \right]+ S'^{(2)}_{\scriptsize \mbox{ghost}}(\alpha)\:.
\label{4}
\end{eqnarray}
Looking at the first line of Eq. (\ref{4}) we find the explicit form
of the operator $\Lambda_\alpha$
\begin{eqnarray}
\Lambda_\alpha = \Delta' -\left(1-\frac{1}{\alpha}\right)d\delta' +
\frac{1}{\alpha} \eta\eta\cdot - \frac{1}{\alpha}(\eta \delta' + d
\eta \cdot)
\label{omega} \:.
\end{eqnarray}
Notice that the use of such an operator is equivalent to employing an
unusual gauge-fixing term in the initial photon Lagrangian which reads
\begin{eqnarray}
\frac{1}{\alpha} \langle  A , ( d -\eta)(\delta'-\eta\cdot)\:
A\rangle\:. \label{fixing}
\end{eqnarray}

\section{The case of the Rindler space}

Let us check the physical results arising from Eq.s (\ref{zeta1'}) and
(\ref{zeta2'}) in the case of the Rindler space. 
Setting $\Omega^2=r^2$ in Eq. (\ref{conformal0}), the related
ultrastatic optical metric reads
\begin{eqnarray}
ds'^2 = d\tau^2 + r^{-2}(dr^2+dy^2+dz^2)
\label{coformalrindler}\:.
\end{eqnarray}
Obviously, this is the natural metric of $S^1\times H^3$ which does
not contain conical singularities. We remind one that $R'^\mu_\nu
=-2\mbox{ diag}(0,1,1,1)$ and $R' = -6$. As for the $1$-form $\eta$
necessary to define the operator $\Lambda_\alpha$, we get
\begin{eqnarray} 
\eta_\mu = \frac{2}{r}\delta^r_\mu \label{etar}\:.
\end{eqnarray}

We want to employ a local $\zeta$-function
\cite{camporesi,libro,report} regularization technique and hence we
define the determinant of an (at least) symmetric operator $L$
through
\begin{eqnarray}
-\frac{1}{2} \ln\det (\mu^{-2} L) = \frac{1}{2}\int d^4x \sqrt{g'}\:
\left[ \zeta'(s=0;x) + \zeta(s=0;x) \ln \mu^2\right],
\label{determinant}
\end{eqnarray}
where the {\em local} $\zeta$ function of the operator $L$ is defined,
as usual, by means of the analytic continuation in the variable $s\in
C$ of the spectral representation of the complex power of the operator
$L$:
\begin{eqnarray}
\zeta(s;x) = \sum_n  \lambda_n^{-s}
 A_n(x) \cdot A_n^\ast(x) \label{zetafunction}\:.
\end{eqnarray}
Above, $A_n(x)$ is a $1$-form eigenfunction of a suitable
self-adjoint extension of the operator $L$ and $\lambda_n$ is its
eigenvalue. The index $n$ stands for all the quantum numbers, discrete
or continuous, needed to specify the spectrum. The set of these modes
is supposed complete and (Dirac, Kroneker) $\delta$ normalized. We
will make also use of the following notation  for the $1$-forms on
$S^1\times H^3$:
\begin{eqnarray}
A \equiv (a | B ) \:, \nonumber
\end{eqnarray}
where $a$ indicates a $1$-form on $S^1$ and $B$ a $1$-form on $H^3$.
All the operations between forms which appear after ``$|$'' are
referred to the manifold $H^3$ and  its metrical structure only. Latin
indices $a,b,c,d,...$ are referred to the coordinates $r,y,z$ on $H^3$
only.

A suitable set of eigenfunctions of the operator $\Delta' -
(1-\alpha^{-1})d\delta'$ as well as $\Lambda_\alpha$ as  can be
constructed using the following complete and normalized set of
eigenfunction of the scalar Hodge de Rham Laplacian on $S^1\times
H^3$:
\begin{eqnarray}
\phi^{({\bf k},n,\omega)}(\tau, r,{\bf x}) = \frac{e^{i{\bf k}{\bf x}}\:
e^{i\nu_n \tau}}
{2\pi^2 \sqrt{\beta}} \sqrt{2\omega \sinh (\pi\omega) }\: r
K_{i\omega}(kr) \label{phi}\:,
\end{eqnarray}
where $\nu_n = \frac{2\pi n}{\beta}$, $n\in Z$, $\omega\in R^+$, ${\bf
k}= (k_y,k_z)\in R^2$, $k= |{\bf k}|$ and all the previous
eigenfunctions have eigenvalue $(\nu^2+ \omega^2 +1)$.
$K_{i\omega}(x)$ is the usual MacDonald function with an imaginary
index. The normalization reads
\begin{eqnarray}
\int d^4x \sqrt{g'}\: \phi^{({\bf k},n,\omega)\ast}\phi^{({\bf
k}',n',\omega')} = \delta^{nn'}\:\delta^2({\bf k}-{\bf
k}')\delta(\omega-\omega')\:.\nonumber
\end{eqnarray}

In the following, we report some relations which are very useful in
checking the results which we shall report shortly. It is convenient
to define the $1$-form $\xi = - d(1/r) = \eta/2r$ on $H^3$. On $H^3$
we have:
 $\delta \eta =4$, $\Delta \eta = 0$, $d \eta= 0$, $d\xi =0$,
 $\Delta \xi = - 3\xi $, $\nabla^{a} \xi_b= - \delta^a_b/r$.
 Furthermore remind that, if $f$ is a 0-form and
$\omega$ an 1-form: 
\begin{eqnarray}
[\Delta (f \omega)]_a = f  [\Delta \omega]_a +
 \omega_a \Delta f - 2 (\nabla_b f) \nabla^b \omega_a\:.
\label{formula1}
\end{eqnarray}
Finally, on a $3$-manifold the following relation holds
\begin{eqnarray}
 \Delta * (\omega\wedge\omega') & = & *[(\Delta \omega)\wedge 
\omega'] +
*[\omega \wedge \Delta \omega'] +R * (\omega\wedge\omega') 
- * [(R \omega)\wedge \omega']\nonumber \\
&& - * (\omega \wedge R \omega') - 2 * (\nabla_a \omega \wedge 
\nabla^{a}\omega')\:,  \label{formula2}
\end{eqnarray}
where obviously $[* (\nabla_a \omega \wedge \nabla^{a}\omega')]_e : 
=\sqrt{g} \epsilon_{ebc} (\nabla_d \omega^{b})\nabla^d \omega'^c$,
$\omega$ and $\omega'$ are $1$-forms and the Ricci tensor acts on
$1$-forms trivially as $(R\omega)_a = R_a^b\omega_b$.
 
\subsection{First optical approach}

Let us now consider the first optical approach, in which we define the
path integral directly in the optical manifold, see Eq.
(\ref{firstcase}). Starting from the scalar eigenfunctions, one can
obtain the following set of eigenfunctions of the operator $\Delta -
(1- \alpha^{-1})d\delta$ on $S^1\times H^3$.
\begin{eqnarray}
A^{(1)} &=& \frac{\sqrt{\omega^2+1}}{|\nu|\sqrt{\omega^2+ \nu^2 +1}}
(\:\partial_\tau \phi \:|\: d\phi \:) =\frac{\sqrt{\omega^2+1}}
{|\nu|\sqrt{\omega^2+ \nu^2 +1}} (\partial_\tau \phi, \partial_r \phi,
\partial_y \phi, \partial_z \phi)
\nonumber\\
A^{(2)} &=& \frac{1}{\sqrt{\omega^2+ \nu^2 +1}} (\:\partial_\tau \phi
\:|\: \frac{-\nu^2}{\omega^2+1} d \phi\:)
\nonumber\\
A^{(3)} &=& \frac{1}{k} (\:0\:|\: *d (\xi \phi) \:) =\frac{1}{k} (0,0,
\partial_z \frac{\phi}{r}, -\partial_y \frac{\phi}{r})
\nonumber \\
A^{(4)} &=& \frac{1}{k\omega}(\:0\:|\: \delta d (\xi \phi)\:)
=\frac{r}{k\omega} (0, \frac{k^2}{r}\phi, \partial_r\partial_y
\frac{\phi}{r}, \partial_r \partial_z \frac{\phi}{r})
\nonumber.
\end{eqnarray}
The last three modes are transverse, $\delta A =0$, whereas the  first
one is a pure gauge mode. From a little Hodge algebra,  the following
normalization relations can be proved:
\begin{eqnarray}
\int d^4x \langle  A^{(J,\omega,n,{\bf k})*}, A^{(J',\omega',n',{\bf
k}')}\rangle =\delta^{JJ'} \delta^{nn'}\:\delta^2({\bf k}-{\bf
k}')\:\delta(\omega-\omega').
\end{eqnarray}
As far as the eigenvalues are concerned, we have:
\begin{eqnarray}
\left[ {\Delta}' - (1- \alpha^{-1}) d \delta' \right] A^{(1)}
&=& \frac{\omega^2 + \nu^2 +1}{\alpha } A^{(1)} \:,\nonumber \\
\left[ {\Delta}' - (1- \alpha^{-1})d \delta'  \right] A^{(2)}
&=& (\omega^2 + \nu^2 +1) A^{(2)} \:,\nonumber \\
\left[ {\Delta}'- (1- \alpha^{-1})d \delta'  \right] A^{(J)}
&=& (\nu^2 + \omega^2) A^{(J)}\hspace{1cm}\mbox{if}\:\:\:\:J = 3,4
\nonumber\:.
\end{eqnarray}
Employing the definition in Eq. (\ref{zetafunction}), the above modes
and the definitions given in the appendix, we have that (notice that
$\phi^\ast$ and $\phi$ take the same values of ${\bf k}$, $n$,
$\omega$)
\begin{eqnarray}
\zeta(s;x) &=&  (\alpha^s+1) \sum_{n=-\infty}^{\infty} \int d^2{\bf k}
d\omega \frac{\phi^*(x) \phi(x) }{[\omega^2 + \nu^2 +1]^s} +
\sum_{n=-\infty}^{\infty} \int d^2{\bf k} d\omega \frac{2 (1+
\omega^{-2})\phi^*(x)\phi(x) }{[\omega^2 + \nu^2]^s} \nonumber\\
&&\nonumber\\
&=&(\alpha^s+1) \zeta^{\scriptsize \mbox{m.c.s.}}(s;x)
+2\zeta^{\scriptsize \mbox{c.c.s.}}(s;x)+
\zeta^{\scriptsize \mbox{extra}}(s;x)
\label{zetafunction'}\:,
\end{eqnarray}
where we have set 
\begin{eqnarray}
\zeta^{\scriptsize \mbox{extra}}(s;x)&=& 2\sum_{n=-\infty}^{\infty}
\int d^2{\bf k} \int \frac{d\omega}{\omega^2} \frac{\phi^*(x)\phi(x)
}{[\omega^2 + \nu^2]^s}\nonumber\\
&=&\frac{\sqrt{\pi}}{\pi^2\beta}\frac{\Gamma(s-\frac{1}{2})}
{\Gamma(s)}\left(\frac{\beta}{2\pi}\right)^{2s-1}\zeta_R(2s-1), 
\label{zetanew}
\end{eqnarray}
so that $\zeta^{\scriptsize \mbox{extra}}(s=0;x)=0$ and
$\zeta'^{\scriptsize \mbox{extra}}(s=0;x)=1/3\beta^2$. Notice that the
second and third terms in Eq. (\ref{zetafunction'}) arise from the
transverse modes $A^{(3)}$ and $A^{(4)}$. The first term in
Eq. (\ref{zetafunction'}) is due to the modes with $J=1,2$.

In calculating Eq. (\ref{zetafunction'}), we encountered Kabat's
surface terms similar to those we encountered in \cite{ielmo}.
However, in the present case all these terms vanish automatically and
no further regularization procedure needs. In fact, all these terms
read as
\begin{eqnarray}
D_r \sum_n \int d{\bf k} \int d \omega 
r^2 K_{i\omega}(kr)K_{i\omega}(kr) f(\omega,\nu,s)\:,\nonumber
\end{eqnarray}
where $D_r$ is an opportune differential operator in $r$. Passing from
the integration variable ${\bf k}$ to the integration variable $r{\bf
k}$, we see that the term after the operator does not depend on $r$,
and so the differentiation produces a vanishing result.

In order to write the complete local $\zeta$ functions of the
electromagnetic field we have to take account of the ghost
contribution. We have already said that in this approach the $\zeta$
function of the ghosts is just minus two times the $\zeta$ function of
a minimally coupled scalar field, but with a gauge-fixing dependent
scale factor, see Eq. (\ref{fantom1}):
\begin{eqnarray}
\zeta_\alpha^{\scriptsize \mbox{ghosts}}(s;x)= -2\zeta^{\scriptsize
\mbox{m.c.s.}}(s;\mu^{-2}\alpha^{-\frac{1}{2}}L_{\xi=0})(x).
\nonumber
\end{eqnarray}
Using this relation, Eq. (\ref{zetafunction'}) and reintroducing
everywhere the renormalization scale $\mu$, we can write the 
complete local $\zeta$ function  of the electromagnetic field as
\begin{eqnarray}
\zeta^{\scriptsize \mbox{em}}(s;x)&=&
(\alpha^s+1) \zeta^{\scriptsize \mbox{m.c.s.}}(s;\mu^{-2}L_{\xi=0})(x)
+2\zeta^{\scriptsize \mbox{c.c.s.}}(s;\mu^{-2}L_{\xi=\frac{1}{6}})(x)
\nonumber\\
&&+\zeta^{\scriptsize \mbox{extra}}(s;\mu^{-2})(x)-2
\zeta^{\scriptsize\mbox{m.c.s.}}
(s;\mu^{-2}\alpha^{-\frac{1}{2}}L_{\xi=0})(x).
\label{zetacompleta}
\end{eqnarray}
It follows that the one-loop effective Lagrangian density is just
\begin{eqnarray}
{\cal{L}}_{\scriptsize \mbox{eff}}(x)&=&\frac{1}{2}
\frac{d}{ds}\left[2\zeta^{\scriptsize \mbox{c.c.s.}}(s;x)+
\zeta^{\scriptsize \mbox{extra}}(s;x)\right]_{s=0}\nonumber\\
&=&\frac{\pi^2}{45\beta^4}+\frac{1}{6\beta^2}.
\label{efflagr}
\end{eqnarray}
We remark the importance of keeping the $\alpha$ dependence of the
action of the ghosts: it gives a contribution proportional to
$\ln\alpha$ which cancels against the ($\ln\alpha$)-dependent term
coming from $(\alpha^s+1)\zeta^{\scriptsize \mbox{m.c.s.}}(s;x)$,
restoring the gauge invariance of the theory. Note also how all the
terms containing $\ln\mu^2$ cancel giving the expected scale invariant
theory.

Integrating this quantity over the manifold and introducing
a cutoff at a distance $\epsilon$ form $r=0$ in order to control
the horizon divergence, we get the one-loop free energy:
\begin{eqnarray}
 F^{(1)}= -\frac{1}{\beta}\int
d^4x\sqrt{g'}{\cal{L}}_{\scriptsize \mbox{eff}}(x)
=-\frac{A_\perp}{1440\pi^2\epsilon^2}
\left[\left(\frac{2\pi}{\beta}\right)^4
+30\left(\frac{2\pi}{\beta}\right)^2\right].
\label{freeen}
\end{eqnarray}
Renormalizing this result in such a way that the internal energy
vanishes at $\beta=2\pi$, we find just the point-splitting result 
\begin{eqnarray}
U_\beta^{(1)\scriptsize \mbox{sub}} = 
U_\beta^{\scriptsize \mbox{phot. p.-s.}} \:.
\end{eqnarray}

\subsection{Second optical approach}

Let us then consider the second optical approach. We were able to
perform the calculations  in the case $\alpha=1$ only, hence a
complete discussion on the gauge invariance ($\alpha$ invariance) is
not possible. However, the found result contains some interest. As
before, the eigenfunctions of the operator $\Lambda_{\alpha=1}$ are
constructed from the scalar eigenfunctions, Eq. (\ref{phi}):
\begin{eqnarray}
A^{(1)} &=& \frac{1}{|\nu|} (\:\partial_\tau \phi \:|\: + \frac{|\nu
|}{2}\eta \phi\:) =\frac{1}{|\nu|} (\partial_\tau \phi, |\nu|
\frac{\phi}{r}, 0, 0)   \nonumber\\
A^{(2)} &=& \frac{1}{|\nu|}(\:\partial_\tau \phi \:|\: - \frac{|\nu
|}{2}\eta \phi\:) = \frac{1}{|\nu|} (\partial_\tau \phi, -|\nu|
\frac{\phi}{r}, 0, 0)   \nonumber\\
A^{(3)} &=& \frac{1}{k} (\:0\:|\: *d (\xi \phi) \:) =\frac{1}{k}(0,0,
\partial_z \frac{\phi}{r}, -\partial_y \frac{\phi}{r}) \nonumber \\
A^{(4)} &=& \frac{1}{k}(\:0\:|\: * (\xi \wedge *d (\xi \phi))\:)
=\frac{1}{k} (0,0, \partial_y \frac{\phi}{r}, \partial_z
\frac{\phi}{r})
\nonumber\:.
\end{eqnarray}
The following normalization relations hold:
\begin{eqnarray}
\int d^4x \langle  A^{(J,\omega,n,{\bf k})*}, A^{(J',\omega',n',{\bf
k}')}\rangle =\delta^{JJ'} \delta^{nn'}\:\delta^2({\bf k}-{\bf
k}')\:\delta(\omega-\omega').
\end{eqnarray}
As far as the eigenvalues are concerned, we have:
\begin{eqnarray}
\Lambda_{\alpha=1} A^{(J)} &=&
 \{ \omega^2 + [(-1)^J + |\nu|]^2 \} A^{(J)} \:\:\:\:\mbox{if}
\:\:\:\: J =1,2\nonumber\:,\\
\Lambda_{\alpha=1} A^{(J)} &=&
(\nu^2 + \omega^2) A^{(J)} \:\:\:\:\mbox{if}\:\:\:\:J = 3,4
\nonumber\:.
\end{eqnarray}
Employing the definition in Eq. (\ref{zetafunction}) and the found
modes we have (notice that $\phi^\ast$ and $\phi$ take the same 
values of ${\bf k}$, $n$, $\omega$)
\begin{eqnarray}
\zeta^{(2)}(s;x) &=&
\sum_{n=1}^{\infty} \int d{\bf k} \int d\omega \frac{2\phi^*(x)
\phi(x) }{[\omega^2 + (\nu+1)^2]^s} +
\sum_{n=1}^{\infty} \int d{\bf k} \int d\omega \frac{2\phi^*(x)
\phi(x) }{[\omega^2 + (\nu-1)^2]^s}  \nonumber\\
&& + \sum_{n=1}^{\infty} \int d{\bf k} \int d\omega \frac{4\phi^*(x)
\phi(x) }{[\omega^2 + \nu^2]^s}  \label{zetafunction2'}\:.
\end{eqnarray}
For simplicity, we have omitted the terms corresponding to $n=0$,
which contribute only to the temperature-independent part of the free
energy: this part will be changed during the renormalization process
(subtraction of the Minkowski vacuum energy). We also stress that
Kabat's surface terms involved during the calculations disappeared
exactly as in the previous approach. The latter term in Eq.
(\ref{zetafunction'}) is due to the modes with $J=3,4$: this term is
exactly twice the $\zeta$ function of a conformally coupled Euclidean
scalar field propagating in $S^1\times H^3$.

As far as the ghost contribution is concerned, it arises from the
action (\ref{5}).  Since the corresponding small fluctuations operator
involves the curvature of the Euclidean Rindler manifold, which has a
Dirac $\delta$ singularity at $r=0$, mathematically it is not well
defined and is not clear how to deal with it. However, as a try we can
suppose to consider $R=0$ and see the consequences.\footnote{See
footnote number 5.} Under this hypothesis, the ghost contribution is
just minus twice that of a conformally coupled scalar field (see Eq.
(\ref{fantom2})) and so it cancels against the contribution of the
modes $J=3,4$.

After having added the ghost contribution, we can write the complete
$\zeta$ function of the electromagnetic field as
\begin{eqnarray}
\zeta^{\scriptsize\mbox{em}}(s;x)
=  \sum_{n=1}^{\infty} \int d{\bf k} \int d\omega \frac{2\phi^*(x)
\phi(x) }{[\omega^2 + (\nu+1)^2]^s}
 +  \sum_{n=1}^{\infty} \int d{\bf k} \int d\omega \frac{2\phi^*(x)
\phi(x) }{[\omega^2 + (\nu-1)^2]^s} \label{zetaphotons2}\:.
\end{eqnarray}
The partition function of the photons is obtained employing the
previous function opportunely continued in the variable $s$ in
Eq. (\ref{determinant}). Dealing with it as in
the previous case, we finally find the free energy
\begin{eqnarray}
F^{(2) \scriptsize \mbox{sub}} = 
-\frac{A_\perp}{1440 \pi^2 \epsilon^2}
\left[ \left( \frac{2\pi}{\beta}\right)^4 - 
30 \left( \frac{2\pi}{\beta}\right)^2
 +29   \right] \:.
\label{freen2}
\end{eqnarray}
In deriving this result we have employed the Riemann zeta function
$\zeta(z, q)$ and its relation with the Bernoulli polynomials
\cite{GR}. This result has the same form as that obtained with the
first approach, Eq. (\ref{freeen}), but the sign in front to the
second term is opposite. The third term is fixed by the
renormalization procedure. The problems arise with  the $\beta^{-2}$
term once again.

In this case it is easy to identify the origin of the discrepancy in our
hypothesis of setting $R=0$ in the ghost action. Nevertheless, there
is some evidence that the origin is not that. In particular, if we
assume that the optical method gives the same results as the
point-splitting one even when $\xi\neq1/6$, then we can suppose
that it is right to substitute the optical result for the ghost
contribution to the above free energy with the point-splitting one
for $\xi=0$. As a result, we get
\begin{eqnarray}
F^{(2) \scriptsize \mbox{sub}} = 
-\frac{A_\perp}{1440 \pi^2 \epsilon^2}
\left[ \left( \frac{2\pi}{\beta}\right)^4 - 
60 \left( \frac{2\pi}{\beta}\right)^2
+59   \right] \:,
\label{freen3}
\end{eqnarray}
which is different from the previous one but still different from
the first optical  approach result. In
particular, the free energy in Eq. (\ref{freen2}) (or (\ref{freen3}))
would yield a negative entropy at the Unruh-Hawking temperature,
which is very hard to  accept on a physical ground.
Summarizing, it seems to us that this second approach, which is the
natural generalization of the procedure used in the scalar case, does
not yield a correct result.

\section{Summary and Discussion} 

The main result of this paper is the proof that the optical method
(the ``first approach'') can be used to compute one-loop quantities in
the Rindler space also in the case of the photon field. The method has
been developed employing a general covariant gauge choice.
Furthermore, by a comparison with other methods, we have seen that
this method produces the same result as the point-splitting procedure.

It is also important to stress that the partition function arising
from our method is completely free from  ``Kabat's'' surface terms.
This is very important because, as we previously said, the approaches
based on the direct computation in the Euclidean Rindler space using
$\zeta$-function or heat-kernel techniques produces such anomalous
terms \cite{kabat,ielmo} and further regularization procedures seem to be
necessary to get physically acceptable results.

We  have also developed a general optical formalism for the Maxwell
field in the covariant gauges based on Hodge de Rham formalism which,
in principle, can be used in different manifolds than the Rindler
space.

However,  many problems remain to be explained. In particular, both in
the photon and in the scalar case the relation between the optical
approach and the direct approach in the manifold with the conical
singularity remains quite obscure. This is due to difficulties
involved in computing the Jacobian of the conformal transformation
in the presence of conical singularities. Moreover, while 
the optical approach can be used in the case of massless
fields without particular difficulties, as soon as the fields have
a mass the optical method becomes much harder to apply. 
In this case, the direct computation in the manifold with 
conical singularities could show its advantages, provided 
one knows how to compute the above Jacobian.

Another general point  which requires further investigation is the
request of self-consistency of the thermodynamics of the gas of
Rindler particle, when the temperature is not the Unruh one. This is a
very important point in calculating the correction to the entropy of a
black hole supposing such corrections due to the fields propagating
around it. We remind one that the Rindler metric approximates the region near
the horizon of a Schwarzschild black hole. The entropy of the fields
is computed using the relation (where $\beta_H$ is the Unruh-Hawking
temperature, $2\pi$ in the Rindler case):
 $S_{\beta_H} = \beta_H^2 \partial_\beta F_\beta |_{\beta_H}$.
In calculating the previous derivative at $\beta=\beta_H$, one has to
consider also the partition function {\em off shell}, namely evaluated
at $\beta \neq 2\pi $ and $\beta$ {\em near} $\beta_H$. It is not so
clear whether it is necessary or not that the thermodynamical laws
hold also for $\beta \neq \beta_H $ and $\beta$ near $\beta_H$ in
order to assure the consistency the procedure followed in calculating
the entropy of the fields at $\beta= \beta_H$. Moreover, it is
well known that the off shell quantum states of a field are affected
by several pathologies on the horizon event.\footnote{Rindler thermal
states with $\beta\neq 2\pi$ violates several axioms of the QFT in
curved backgrounds. For example, see \cite{haaglibro} and ref.s
therein.} Furthermore, they are unstable states in a semiclassical
approach to
quantum gravity due to the divergence of the renormalized stress
tensor on the horizon. Thus, it is reasonable to wonder about
the thermodynamical consistency of the  results when one works
off shell. We conclude with a discussion on this point.

Let us first consider the point-splitting result and the annoying
dependence on the parameter $\xi$ of the massless scalar field results
and its relation with the request of a consistent thermodynamics. As
we have already said, on more regular manifolds the (integrated)
physical quantities should  not depend on the actual value of $\xi$,
whereas in the case of Euclidean Rindler wedge the conical singularity
introduces an, apparently unphysical, $\xi$ dependence in the physical
quantities.\footnote{Notice that the case of the cosmic string theory
is quite different because different values of $\xi$ correspond to
different internal structures of the string. This is obvious by
considering a string with a finite thickness,  which has a nonvanishing 
curvature within itself. In the limit of a vanishing
thickness, the curvature $R$ gets a Dirac $\delta$ behaviour along the
string in the Lorentzian manifold.}
A similar problem occurs even in flat spaces in the presence of boundaries
\cite{birrel}. In those cases, one can see that the renormalized
energy-momentum tensor diverges on the boundary, unless the coupling
is conformal. So, one could say that the conformal coupling is, in
this sense, more ``physical'' than the others. By this we mean that it
behaves like a real field, such as the Maxwell field.

Inspired by this fact, we shall look for a criterion to choose a value 
of $\xi$ which is more ``physical'' than the others as far as
the thermodynamics is concerned. 
Thus we shall discuss the consistency of the thermodynamics of the
point-splitting results.

{}From the thermodynamics, we know that we can obtain the internal
energy $U_{\beta,\xi}^{\scriptsize \mbox{p.-s.}}$ (see Eq. (\ref{i3}))
as the derivative with respect to $\beta$ of an appropriate free 
energy $F_{\beta,\xi}^{\scriptsize \mbox{p.-s.}}$ multiplied by
$\beta$, possibly corrected by an suitable energy-subtraction
procedure. Taking  the space homogeneity along the $y$ and $z$
directions into account, we found the form of the free energy
\begin{eqnarray}
F_{\beta,\xi}^{\scriptsize \mbox{p.-s.}} &=&
-\frac{L_y L_z}{2880\pi^2\epsilon^2} \left[
\left(\frac{2\pi}{\beta}\right)^4 - 30 (6\xi -1)
\left(\frac{2\pi}{\beta}\right)^2 - 30 (6\xi -1) +3 \right]  
\nonumber\\
& & - \frac{L_y L_z}{2880\pi^2\epsilon^2} \left[ {\cal
U}(\xi,\epsilon) +\frac{f(\xi, \epsilon)}{\beta} \right]
\label{i4}\:.
\end{eqnarray}
The unknown function $f(\xi,\epsilon)$ can be dropped by requiring
that the entropy $S_{\beta,\xi} = \beta^2 \partial_\beta
F_{\beta,\xi}^{\scriptsize \mbox{p.-s.}}$ vanishes at $\beta
\rightarrow + \infty$. The function ${\cal U}$, which does not depend
on $\beta$ but can depend on the geometry background, is necessary 
due to the fact that the energy in Eq. (\ref{i3}) is the
Minkowski renormalized one, but we want to remain on a more general
ground in order to use the thermodynamical laws. In other words, we
may notice that the energy in Eq. (\ref{i3}) becomes negative if the
temperature is sufficiently low, for example,
in the most interesting  range $0 \leq \xi \leq 1/6$, and hence such
an energy cannot directly arise form a statistical partition  function
but a further subtraction procedure must have taken place. The
function ${\cal U}$ takes into account this energy subtraction
procedure.

{}From statistical thermodynamical laws, one expects that the
$y$ and $z$ principal pressure, namely $T_{yy}$ and $T_{zz}$
in Eq. (\ref{tensor}), integrated over $dz dr \sqrt{g}$ and $dy dr
\sqrt{g}$ respectively, can be obtained  taking the  $L_y$
($L_z$) derivative of the previous free energy, with the sign changed
and ${\cal U}$ opportunely chosen. An easy computation shows  that,
due to the terms containing $\beta^{-2}$, this does not hold for any
value of $\xi$, but only in the conformally coupled case, $\xi =1/6$.
After the Minkowskian energy subtraction, the corresponding free
energy reads
\begin{eqnarray}
F_\beta^{\scriptsize \mbox{p.s.}} 
:= F_{\beta,  \xi=1/6}^{\scriptsize \mbox{p.-s. sub}} =
-\frac{A_\perp}{2880\pi^2\epsilon^2}\left[
\left(\frac{2\pi}{\beta}\right)^4 +3 \right].
\label{i5}
\end{eqnarray}
This is just the free energy obtained by the optical method after the 
Minkowski renormalization. Therefore, it seem that only in the 
conformally coupled case the stress tensor (\ref{tensor}) yields a 
consistent thermodynamics, at least 
 as far as the relation between energy and pressures
 is concerned.

Now, let us consider the photon case. In such a case we have not the
freedom to adjust a parameter in the stress tensor in order to agree
with the thermodynamics. The free energy we find from the total 
energy in the case of the photon stress tensor of Eq. (\ref{tensor2}) 
reads:
\begin{eqnarray}
F_{\beta}^{\scriptsize \mbox{phot. p.-s. }}&=&
 -\frac{L_y L_z}{1440 \pi^2 \epsilon^2}\left[
\left(\frac{2\pi}{\beta}\right)^4 +30\left(
\frac{2\pi}{\beta}\right)^2 - 33\right]\nonumber\\
& & -\frac{L_y L_z}{1440 \pi^2 \epsilon^2}\left[{\cal U}(\epsilon) +
\frac{f(\epsilon)}{\beta} \right].
\label{Fpsphot}
\end{eqnarray}
As before, we can drop the term containing the undetermined function
$f(\epsilon)$ by requiring a vanishing entropy in the limit
$\beta\rightarrow +\infty$. The above free energy produces the
point-splitting internal energy and, after the Minkowski
renormalization, it coincides with the free energy obtained by
renormalizing that obtained by the optical approach, Eq.
(\ref{freeen}).

The point is that if we apply the above procedure to compute the
integrated principal pressures along the  $y$ and $z$ directions to
the above photon free energy, there is no way to choose ${\cal U}$ in
such a way to get the same result as integrating the  $yy$ and $zz$
components of the photon stress tensor in Eq. (\ref{tensor2}). This is
due to the presence of a term proportional to $\beta^{-2}$ and the
independence on $\beta$ of the function ${\cal U}$.

In order to get the ``correct'' pressures (but a wrong internal energy!)
employing the derivatives as previously pointed out, one should
take a free energy which is twice that in Eq. (\ref{i4}) with 
$\xi= 1/9$, $f=0$ and  ${\cal U}$ opportunely chosen.

Hence, it seems that the point-splitting stress tensor of photons in
the Rindler wedge does not give a consistent
thermodynamics.\footnote{This problem arises also dealing with the
massless spinorial field as it simply follows from the point-splitting
renormalized stress tensor obtained in \cite{FRSE87} (analytically
continued from the cosmic string   to the Rindler space).} It is very
important to remark that the above thermodynamical argument cannot
be applied to the cosmic string theory, since in that case the stress
tensor in Eq. (\ref{tensor2}) is the zero-temperature one, and $\beta$
is not the inverse of the temperature.

In a pessimistic view, this problem and the $\xi$ dependence of the
integrated quantities in the scalar case could be considered as
another proof of the inconsistency of the Rindler theory (and maybe of
the Schwarzschild theory) when one works at temperatures different
from the Unruh-Hawking one, and a discouraging result for the
attempt to evaluate the correction to the Bekenstein-Hawking entropy
through the ``off shell'' procedure.\footnote{However, it
could be possible to interpret the entropy formula
at Hawking's temperature, without making use of thermodynamical laws
off shell. Maybe,  possible ways could arise studying the {\em
geometrical} entropy   employing the replica trick
\cite{CW94,LAWI96}.}

\par \section*{Acknowledgments}
We are grateful to  Francesco Belgiorno, Guido Cognola, Giuseppe
Nardelli,  and in particular to Giampiero Esposito, Marco Toller,
Luciano Vanzo,  Sergio Zerbini, for valuable discussions and useful
suggestions.

\section{Appendix} 

In computing the photon $\zeta$ function on $S^1\times H^3$ one meets
the $\zeta$ function of a scalar field in the same background, both in
conformal and minimal coupling. Therefore, it is useful to report here
these  $\zeta$ functions. The small fluctuations operator for a scalar
field in the optical metric is
\begin{eqnarray}
L_\xi=\Delta-6\xi=-[\partial_\tau^2-r\partial_r
+r^2\partial_r^2+6\xi].
\nonumber
\end{eqnarray}
where $\Delta$ is the Hodge de Rham Laplacian on $S^1\times H^3$. A
complete set of eigenfunctions has been given in the main text,
Eq. (\ref{phi}), with eigenvalue $[\nu_n^2+\omega^2+1-6\xi]$.
Therefore, the local $\zeta$ function is
\begin{eqnarray}
\zeta(s|L_\xi)(x)&=&\sum_{n=-\infty}^\infty
\int_0^\infty d\omega\int d^2{\bf k}\,[\nu_n^2+\omega^2+1-6\xi]^{-s}
\phi^\ast(x)\phi(x)\nonumber\\
&=&\frac{\sqrt{\pi}}{8\pi^2\beta}\frac{\Gamma(s-\frac{3}{2})}
{\Gamma(s)}\sum_{n=-\infty}^\infty
\int_0^\infty d\omega\,\omega^2[\nu_n^2+\omega^2+1-6\xi]^{-s}
\nonumber\\
&=&\frac{\sqrt{\pi}}{8\pi^2\beta}\frac{\Gamma(s-\frac{3}{2})}
{\Gamma(s)}\left(\frac{2\pi}{\beta}\right)^{3-2s}
\left[2E\left(s-\frac{3}{2};\frac{\beta}{2\pi}\sqrt{1-6\xi}\right)-
\left(\frac{\beta}{2\pi}\sqrt{1-6\xi}\right)^{3-2s}\right],
\nonumber
\end{eqnarray}
where $E(s;a)=\sum_{n=0}^\infty [n^2+a^2]^{-s}$ is the Epstein
$\zeta$ function. In the conformally coupled case, the Epstein 
function becomes a Riemann $\zeta$ function and so
\begin{eqnarray}
\zeta^{\scriptsize \mbox{c.c.s.}}(s;x)\equiv
\zeta(s|L_{\xi=\frac{1}{6}})(x)=\frac{\sqrt{\pi}}{4\pi^2\beta}
\left(\frac{\beta}{2\pi}\right)^{2s-3}\frac{\Gamma(s-\frac{3}{2})}
{\Gamma(s)}\zeta_R(2s-3),\nonumber
\end{eqnarray}
One can easily check that 
$\zeta^{\scriptsize \mbox{c.c.s.}}(s;x)|_{s=0}=0$ and
$$
\frac{d}{ds}\zeta^{\scriptsize
\mbox{c.c.s.}}(s;x)|_{s=0}=\frac{\pi^2}{45\beta^4}. $$
Another important case is the minimally coupled one, $\xi=0$, for
which there is not a more explicit form. However, using the identity
$$
E(s;a)=\frac{1}{2a^{2s}}+
\frac{\sqrt{\pi}}{2}\frac{\Gamma(s-\frac{1}{2})}{\Gamma(s)}a^{1-2s}
+\frac{2\sqrt{\pi}}{\Gamma(s)}\sum_{n=1}^\infty (\frac{\pi
n}{a})^{s-\frac{1}{2}} K_{s-\frac{1}{2}}(2\pi n a).
$$
and the  fact that the MacDonald function $K_\nu(x)$ is analytic in
the index $\nu$ and decays exponentially as $|x|\rightarrow\infty$
so that the third term in the previous expansion is analytic in $s$
(and vanishes as $s \rightarrow 0$),
we find that the $\zeta$ function does not vanish in 
$s=0$:
$$\zeta^{\scriptsize \mbox{m.c.s.}}(s;x)|_{s=0}=\frac{1}{32\pi^2}.$$
We do not know the value in zero of the derivative, but it is not 
required in our computations.

\end{document}